\documentclass[a4paper, 11pt]{article}

\usepackage{amssymb}
\usepackage{graphicx}
\usepackage{epstopdf}
\usepackage{amsmath}
\usepackage{hyperref}
\usepackage{color}
\usepackage{xcolor}
\usepackage{listings}

\definecolor{gray}{rgb}{0.4,0.4,0.4}
\definecolor{darkblue}{rgb}{0.0,0.0,0.6}
\definecolor{cyan}{rgb}{0.0,0.6,0.6}

\lstdefinestyle{listXML}{language=XML, extendedchars=true,  belowcaptionskip=5pt, xleftmargin=1.8em, xrightmargin=0.5em, numbers=left, numberstyle=\small\ttfamily\bf, 
basicstyle=small
frame=single, breaklines=true, breakatwhitespace=true, breakindent=0pt, emph={}, emphstyle=\color{red}, basicstyle=\small\ttfamily,
columns=fullflexible, showstringspaces=false, commentstyle=\color{gray}\upshape,
morestring=[b]",
morecomment=[s]{<?}{?>},
morecomment=[s][\color{orange}]{<!--}{-->},
keywordstyle=\color{cyan},
stringstyle=\color{black},
tagstyle=\color{darkblue},
morekeywords={xmlns,version,type}
}

\title{New model for datasets citation and extraction reproducibility in VAMDC}

\author{Carlo Maria Zw\"olf\thanks{LERMA, Observatoire de Paris, PSL Research University,  CNRS, 
Sorbonne University, UPMC Univ Paris 06,  5 Place Janssen, 92190 Meudon, France}, Nicolas Moreau\footnotemark[1], 
and Marie-Lise Dubernet\footnotemark[1]}

\begin{document} 
\maketitle

\begin{abstract}
In this paper we present a new paradigm for the identification of datasets extracted from the
Virtual Atomic and Molecular Data Centre (VAMDC) e-science infrastructure. Such identification includes
information on the origin and version of the datasets, references associated to individual data in the datasets,
as well as timestamps linked  to the extraction procedure. This paradigm is described through the modifications
of the language used to exchange data within the VAMDC and through the services that will implement those
modifications. This new paradigm should enforce traceability of datasets, favour reproducibility of datasets extraction,
and facilitate the systematic citation of the authors having originally measured and/or calculated the extracted atomic and molecular
data.
\end{abstract}


\section{Introduction}
\label{intro}
Atomic and Molecular Data Providers are often at the forefront in defining new paradigms for the dissemination of their data to the scientific community:\\

Since the second half of 1990s, with the spread of the Internet, some historical databases such as VALD \cite{vald1}, HITRAN \cite{hitran83}, \cite{hitran1}, the Submillimeter, millimeter and microwave spectral line catalog (so-called JPL Catalogue) \cite{jpl}, CDMS \cite{cdms01}, \cite{cdms1}, GEISA \cite{geisa98}, \cite{geisa} were precursor of a free online sharing of data, thus anticipating of some decades trends later generalised by gouvernements and institutions with the {\it open data} and {\it open knowledge  initiatives} \cite{Huji2011}, \cite{Molloy}, \cite{Piwowar}. \\

During the last ten years, Atomic and Molecular resources were highly fragmented and available through a variety of interfaces. This situation negatively impacted the full exploitation of the scientific data and has been a major bottleneck between the data providers and the wide body of users. 
From 2009 to 2014, the goal of the European FP7 projects {\it VAMDC} \cite{vamdc10} and its successor {\it SUP@VAMDC} \cite{zwolf2014} was to provide the  international community with a unique technical and collaborative framework for Atomic and Molecular data sharing: the VAMDC International Consortium \cite{dub2016}. The VAMDC technical e-infrastructure federates, in an interoperable way, a wide range of heterogeneous databases containing data coming from different physical communities, provides a single entry-point both for discovering resources and for accessing the available data in a unified  way (http://portal.vamdc.eu). For building and maintaining this e-infrastructure, VAMDC has developed specific procedures, protocols, data formats and query languages (http://standards.vamdc.eu).\\
The Atomic and Molecular data providers community, organised around the VAMDC Consortium, has thus anticipated the identification and resolution of problems that are currently discussed in international groups belonging to the Research Data Alliance (RDA).  The goal of RDA is indeed to build the social and technical bridges that enable open sharing of data in the current highly fragmented global research data landscape. \\

Using the VAMDC facilities, scientists can now easily discover the atomic and molecular resources, and can access in a unique and practical way their contained data. However, the adoption of VAMDC by a large community has revealed a new set of challenging issues. The VAMDC infrastructure data are dynamic. A database exposed through the VAMDC infrastructure may evolve over time: the most recent and precise version of  given data may replace old ones as in VALD \cite{vald1, vald2}, or new datasets may be added  to the already existing ones as in BASECOL \cite{basecol1, basecol2}.  It may happen that some of these database evolutions are not systematically reported  through new publications. \\

Assume then that a scientist extracts from VAMDC at a given time a ``dataset'', composed of an ensemble of ``data'',  and wishes to use this ``dataset'' in order to produce some science that will be published  into a scientific paper: how can he/she cite this ``dataset'' and the individual ``data''? Since the database content may evolve, for the consistency of the scientific publication, the citation should refer to ``datasets'' well defined in space (where the ``dataset'' physically comes from) and time (at what time the ``dataset'' was produced and extracted). In addition the citations should contain pointers to the authors who originally measured, calculated and/or fitted the individual data. Moreover, for the reproducibility of the scientific process described into the paper referencing the ``dataset'', everybody wishing to verify step-by step the procedures described into the paper, should be able to easily recover the original ``dataset'' and replay the data-production workflow.

In spring 2014 we joined the RDA Data Citation Working Group in order to study these open questions. The goal of this Working Group \cite{rdawgdc-recdocument} is to create the identification mechanisms that 
\begin{itemize}
\item allows us to identify and cite arbitrary views of a ``dataset'' in a precise, machine actionable manner;
\item allows us to cite and retrieve that ``dataset'' as it existed at a certain point in time, whenever the database is static or highly dynamic;
\item is stable across different technologies and technological changes. 
\end{itemize}

The very new model we will discuss through this paper is the mechanisms we may apply on the data coming from the VAMDC e-infrastructure in order to meet the recommendations of the RDA Working Group on Data Citation with the goal of obtaining
\begin{itemize}
\item a rigorous and reproducible data citation model,
\item the reproducibility of data production (or data processing) workflows, 
\end{itemize} 
and in order to facilitate the citation of the original authors of the atomic and molecular data.

\begin{figure}[htbp]
\begin{center}
\includegraphics[width=0.4\textwidth]{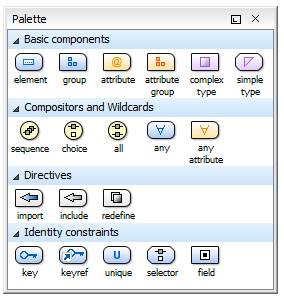} 
\caption{Graphical convention adopted for building graphical representations, starting from the XML schema.}
\label{figSchema}
\end{center}
\end{figure}

\begin{figure}[htbp]
\begin{center}
\includegraphics[width=1.0\textwidth]{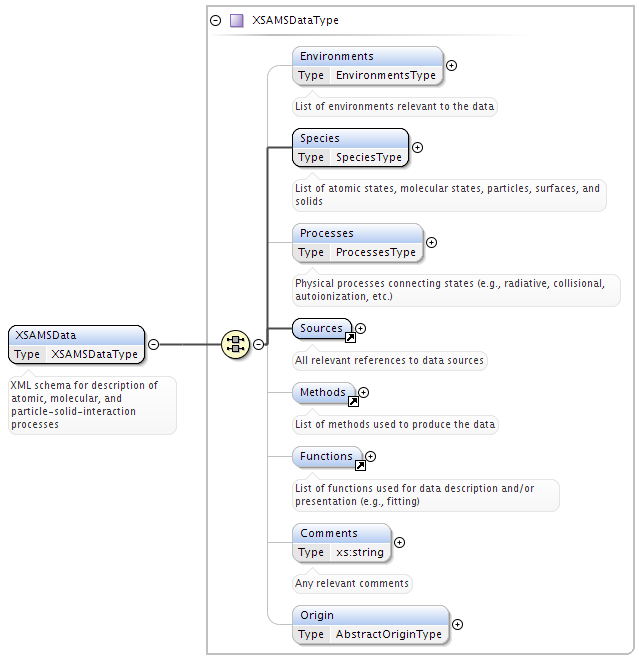} 
\caption{{Representation of the root element of an XSAMS file.}}
\label{rootXsams}
\end{center}
\end{figure}

\section{Existing VAMDC solution for data citation}\label{existingSolution}
All the output of the VAMDC infrastructure are built upon an abstract object model: each specific output is an instance of this model. For historical reasons, without loss of generality, the model has been fixed into an XML format, the XSAMS XSD schema\footnote{XSAMS is the acronym for: XML Schema for Atoms, Molecules and Solids}.\\

\subsection{Graphical convention for representing XML elements}
The graphical diagrams displayed in this document are a simple rendering of every XML element contained in the XSAMS, and they are obtained following the graphical conventions of  Fig.~\ref{figSchema}. In this graphical representation (see Fig.~\ref{rootXsams} for example)  every complex element is linked with segments to the contained sub-elements. A bold segment indicates that the sub-element is required and a thin segment indicates that the sub-element is optional. Moreover, the cardinality of the contained sub-elements can be expressed on the segments. The detailed explanation of the XSAMS components can be found in the official VAMDC XSAMS documentation at http://www.vamdc.org/standards. \\

\subsection{Example of data-citation need for scientific authors}
\label{scienceNeeds}
Surveys of interstellar regions requires the use of spectroscopic information within the observed range of wavelengths/frequencies. As an example, the survey by \cite{astro} covers frequencies from $83302$ MHz to $262404$ MHz and detect emission from about  36 species. For that survey, \cite{astro} indicate that they used catalogues from two public databases \cite{jpl}, \cite{cdms1} and one private database of J. Cernicharo (private communication). We note that there is no knowledge of the exact dataset used in the analysis,  and therefore the analysis may not be reproductible if the database contents evolve over the years. Secondly, we note there is no citation of the authors who produced the spectroscopic data. Obviously for such large surveys with so many species there is a large contribution from many experimental/theoretical spectroscopic papers. On the contrary, that for the non-local thermodynamical equilibrium analysis of spectra (that includes the use of collisional rate coefficients) about 12 publications related to collisional data are cited. 
This dichotomy of treatment could be first explained by the complexity of citing/finding many spectroscopy authors, while it is easy to cite a few collisional papers. \\
Similarly, another survey \cite{corby2015} cites many spectroscopic databases without
citing either the original authors or the version of data used in the survey's analysis. 
The study of Punanova et al. \cite{punanova16} cites the authors of transitions that are not part of a database, such
as the hyperfine transitions of N$_2$H$^+$ \cite{pagani09} and such as the $1 \rightarrow 0$ transition 
of C$^{17}$O \cite{frerking81}, but  they
cite the splatalogue catalog (http://www.cv.nrao.edu/php/splat/) for the $1 \rightarrow 0$ transition of C$^{17}$O. Those
examples confirm that the origin and versioning of spectroscopic data might be lost along the chain of usage as
soon as spectroscopic data are taken from a database.\\

\begin{figure}[htbp]
\begin{center}
\includegraphics[width=1.0\textwidth]{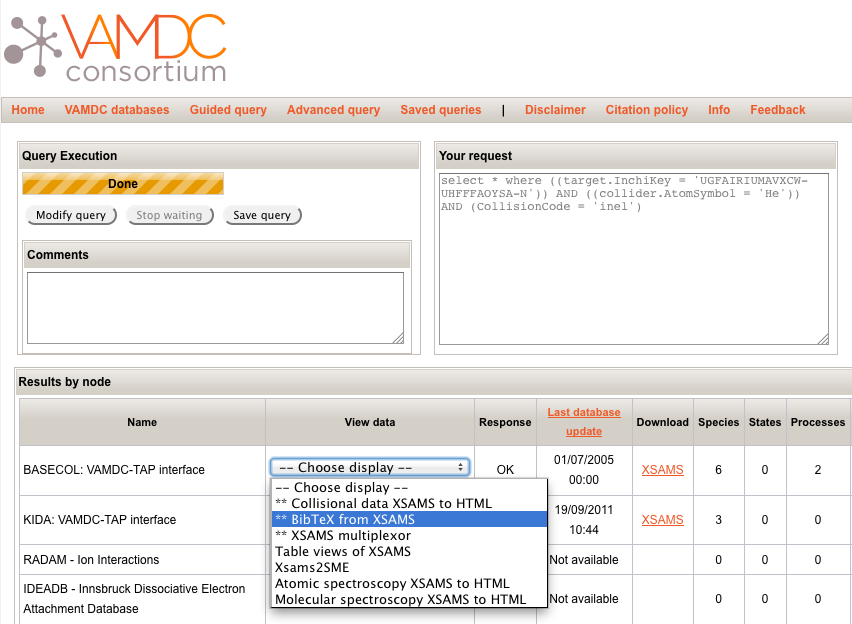} 
\caption{{ Detail of the result page of the VAMDC portal highlighting the ``bibtex'' processor.}}
\label{portalRes}
\end{center}
\end{figure}

\begin{figure}[htbp]
\begin{center}
\includegraphics[width=0.6\textwidth]{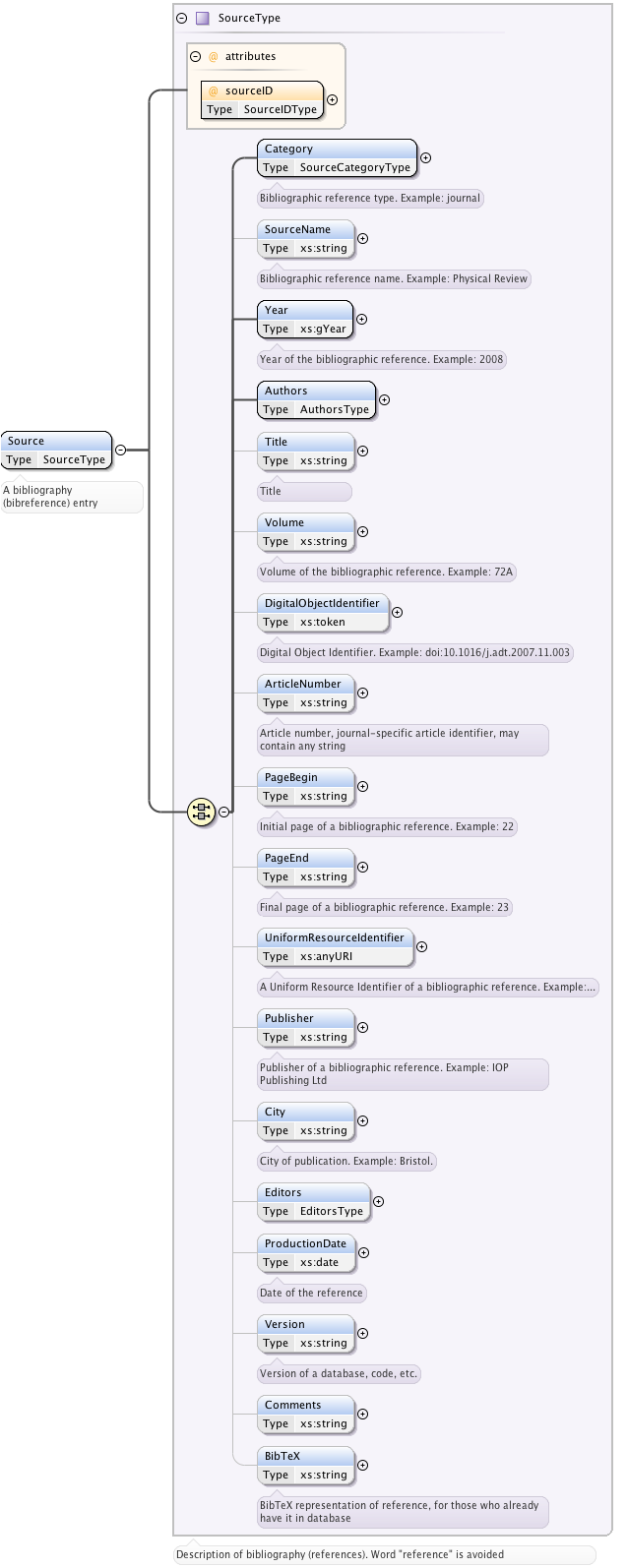} 
\caption{{ Details of the {\it Source} field contained in XSAMS files.}}
\label{sourcesXSAMS}
\end{center}
\end{figure}

Through the VAMDC e-infrastructure, the protocols presently allow to retrieve the references associated to the spectroscopic and collisional data in a straightforward and quick manner. For example, from the result page of the VAMDC portal (cf. figure \ref{portalRes}) the ``bibtex'' visualisation processor allows one to get references attached to the data in a usable form for publication. This processor produces XSAMS output file (produced by the VAMDC infrastructure) by extracting the bibliographic information contained into the {\it Sources} field, which is embedded into each data entry (cf. figure \ref{rootXsams} and \ref{sourcesXSAMS}).
Thus, the use of the current version of VAMDC allows full citation of any data and such citation relies on the good will of the users. Even so, the citation process may become cumbersome when the extracted datasets come from many sources. \\

This latter issue and the former issue of reproducibility of the extraction process are addressed in the new paradigm of datasets citation that we describe below.  

\section{Introducing the evolutions of the XSAMS format}\label{xsamsEvol}
In this paragraph we discuss new mechanisms allowing one to identify through time and location, in a sustainable and unique way, the data sets extracted by users from the VAMDC e-infrastructure. The proposed mechanisms comes with an evolution of the XSAMS standard.\\
The practical usage of the described features is presented in section \ref{queryStore}

\subsection{Introduction of the new schema elements for handling data versioning and origin}

The root element of a XSAMS file is the {\it XSAMSData} object (cf. figure \ref{rootXsams}), which is an instance of the {\it XSAMSDataType} class. The innovation compared with the current official version of the XSAMS format (version 12.07) is the introduction of the {\it Origin} field, that contains the new mechanisms for handling the datasets citations.\\

\begin{figure}[htbp]
\begin{center}
\includegraphics[width=1.0\textwidth]{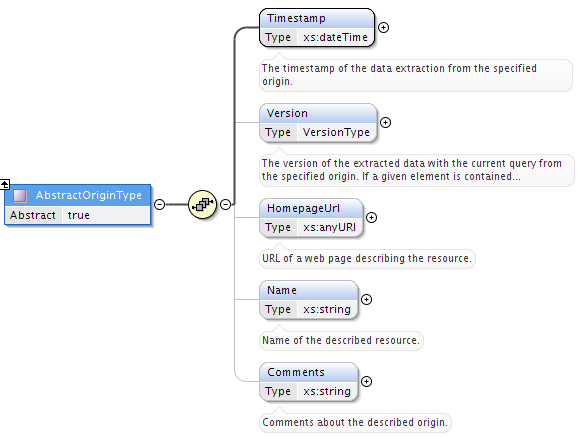} 
\caption{{ Representation of the abstract element {\it AbstractOriginType}.}}
\label{abstractOrigin}
\end{center}
\end{figure}

The {\it Origin} field is an instance of a class extending the {\it AbstractOriginType} (cf. figure \ref{abstractOrigin}). This abstract type must contain:
\begin{itemize}
\item a unique {Timestamp}. This field defines the exact time when the data have been extracted by the infrastructure;
\item a {\it Version} field, instance of {\it VersionType} class (cf. paragraph \ref{versionElement});
\item a {\it HomepageURL} field, which is the link to the website describing the resource that produced the data;
\item a {\it Name} field, which is the name of the resource producing the data;
\item a {\it Comments} field, which (as its name indicates), contains the comments related with the resources producing the data. 
\end{itemize}
The three classes extending the abstract type {\it AbstractOriginType} are: {\it VamdcNodeOriginType}, {\it VamdcProcessorOriginType}, and {\it OtherOriginType}.\\

\begin{figure}[htbp]
\begin{center}
\includegraphics[width=1.0\textwidth]{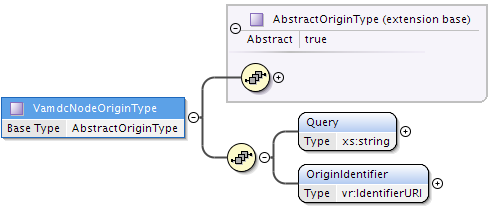} 
\caption{{ Representation of the element {\it VamdcNodeOriginType}.}}
\label{nodeOrigin}
\end{center}
\end{figure}

The {\it VamdcNodeOriginType} (cf. figure \ref{nodeOrigin}), as its name indicates, is used when the resource producing the data is a VAMDC node. It extends the {\it AbstractOriginType} by mandatorily  adding:
\begin{itemize}
\item a {\it Query} field, which is a text string defining the query used for extracting the data from the described node. 
\item an {\it OriginIdentifier} field, which is the identifier of the node into the VAMDC infrastructure, as it appears into the VAMDC registry. \\
\end{itemize}

\begin{figure}[htbp]
\begin{center}
\includegraphics[width=1.0\textwidth]{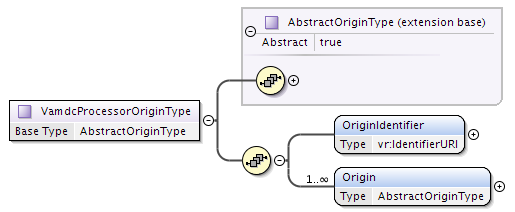} 
\caption{{Representation of the element {\it VamdcProcessorOriginType}.}}
\label{processorOrigin}
\end{center}
\end{figure}

The {\it VamdcProcessorOriginType} (cf. figure \ref{processorOrigin}), as its name indicates, is used when the resource producing the data is a VAMDC processor. It extends the {\it AbstractOriginType} by mandatorily adding:
\begin{itemize}
\item an {\it OriginIdentifier} field, which is the identifier of the processor into the VAMDC infrastructure, as it appears into the VAMDC registry. \\
\item the list (which must contain at least one element) of the {\it Origins} that produced the data used as inputs by the current processor. 
\end{itemize}

\begin{figure}[htbp]
\begin{center}
\includegraphics[width=1.0\textwidth]{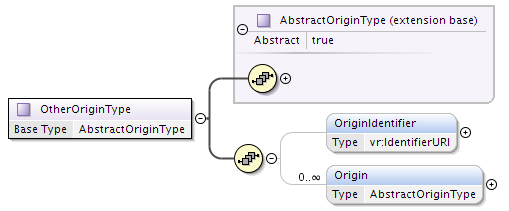} 
\caption{{Representation of the element {\it OtherOriginType}.}}
\label{otherOrigin}
\end{center}
\end{figure}

The {\it OtherOriginType} (cf. figure \ref{otherOrigin}), as its name indicates, is used when the resource producing the data is not a VAMDC node and neither a VAMDC processor. It extends the {\it AbstractOriginType} by  adding:
\begin{itemize}
\item an {\it OriginIdentifier} field, which is the identifier of the resource into the VAMDC infrastructure, as it appears in the VAMDC registry. \\
\item the list of the {\it Origins} that produced the data used as inputs by the current resource. 
\end{itemize}

\begin{figure}[htbp]
\begin{center}
\includegraphics[width=1.0\textwidth]{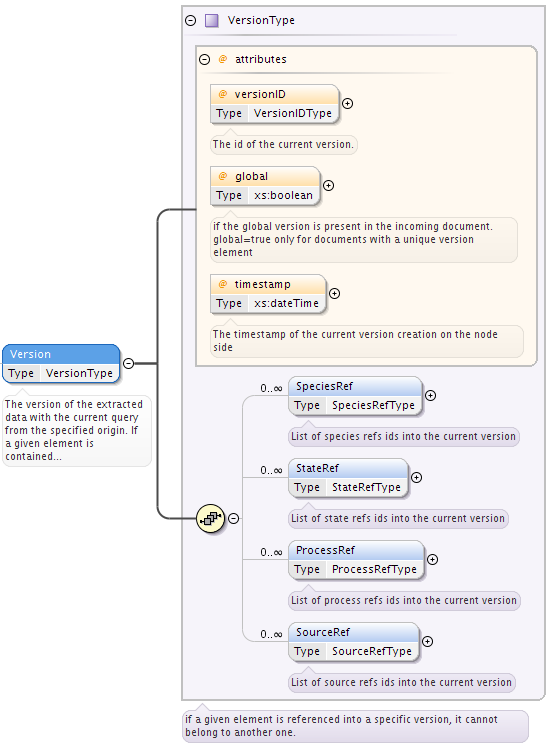} 
\caption{{Representation of the element {\it VersionType}.}}
\label{versionType}
\end{center}
\end{figure}

These three classes make sure that the users know what is the latest treatment of the data, and what are the original sources of the
data, thus making possible to trace the original public exposure of the data. 

\subsubsection{The Version element}\label{versionElement}
With the  {\it VersionType} class the data contained in a XSAMS file are managed with fine grained granularity (cf. figure \ref{versionType}). 
This class contains three attributes:
\begin{itemize}
\item a {\it versionID} which is the unique identifier of the current version instance;
\item a {\it global} boolean field. If global is true, then XSAMS file should consist of a unique instance of {\it VersionType} and all the data contained into the file will automatically belong to this Version. This implies that the attribute {\it global} equal to true is legal for XSAMS document with a unique {\it Origin} element (since {\it Origin} automatically contains a {\it Version}).
\item a unique {\it timestamp} field which define the timestamp of the current version creation on the resources providing the data.
\end{itemize}
Moreover, this class contains three lists of elements:
\begin{itemize}
\item a list of {\it SpeciesRef} objects of type {\it SpeciesRefType}. In other words, this list contains the references to all the species (defined elsewhere into the XSAMS file) which belong to the current Version;
\item a list of {\it StateRef} object of type {\it StateRefType}. This list contains the references to all the states (defined elsewhere into the XSAMS file) which belong to the current Version;
\item a list of {\it ProcessRef} object of type {\it ProcessRefType}. This list contains the references to all the process (defined elsewhere into the XSAMS file) which belong to the current Version.
\item a list of {\it SourceRef} object of type {SourceRefType}. This list contains the references to all the bibliography sources (defined elsewhere into the XSAMS file) which belong to the current version.
\end{itemize}
If a given element is referenced into a specific {\it Version} object, it can not belong to another one. \\

{\bf Remark:} It is important to underline the difference between the {\it timestamp} field contained into an {\it Origin} object and the {\it timestamp} contained into a {\it Version} object. The first corresponds to the time when the data were extracted by the user (i.e. the data when the file was created) while the second corresponds to the time when that specific version of data was made available to community by the resource providing the data.

\section{Some examples of data extracted from VAMDC, formatted with the proposed schema}\label{xsamsExamples}
In this paragraph we will provide examples of data extracted from the VAMDC infrastructure and formatted with the schema described in the previous section. 
For reason of clarity, we have lightened those examples, by manually selecting only a subset of the data really available into VAMDC. This will not affect the methods and the accuracy of the described approach.\\

The VAMDC internal convention  is to use wavelength expressed in angstrom when querying radiative data from databases. This is imposed by the need of interoperability within VAMDC. Nevertheless from our portal  (\cite{dub2016}, \hyperref[http://portal.vamdc.eu]{http://portal.vamdc.eu}) the users may query the radiative databases using whatever relevant quantities (frequency, energy, wavenumber, wavelength) associated to a choice of relevant units. No change in the original data is performed in the extracted output. For example \ref{cdmsFile}, line 108 shows that CDMS provides frequencies in MHz.

\subsection{Example of an extraction from the BASECOL node}
Let us consider an elastic collisional process, where the target is $CO$ and the collider is $He$. Let us try to find some data on this collision into the VAMDC infrastructure. The associated query (generated by the portal from the Graphical User Interface and passed to all the VAMDC nodes) is: \\ \\
{\tt select * where \\((target.MoleculeStoichiometricFormula = 'CO')) AND \\((collider.AtomSymbol = 'he'))}\\

The file reported in its entirety in \ref{besecolFile} is the output that the BASECOL node \cite{basecol2} generates when answering the previous query (with the reservations made at the beginning of paragraph \ref{xsamsExamples}). Let us browse the content of this file and focus on the {\it Origin} element:
\begin{lstlisting}[style=listXML]
   <Origin xsi:type="VamdcNodeOriginType">
    <Timestamp>2015-12-03T14:40:21+01:00</Timestamp>
    <Version versionID="VER001" global="false" timestamp="2015-09-01T08:10:12+01:00">
      <SpeciesRef>XBAS2</SpeciesRef>
      <SpeciesRef>XBAS52</SpeciesRef>
      <StateRef>SBASET54-1</StateRef>
      <StateRef>SBASET52-1</StateRef>
      <StateRef>SBASET52-2</StateRef>
      <ProcessRef>PBASC50t2T1c1C1</ProcessRef>
      <SourceRef>BBAS0</SourceRef>
      <SourceRef>BBAS849</SourceRef>
    </Version>
    <HomepageUrl>http://basecol.vamdc.org</HomepageUrl>
    <Name>Basecol</Name>
    <Query>select * where ((target.MoleculeStoichiometricFormula = 'CO')) AND 
    ((collider.AtomSymbol = 'he'))</Query>    
    <OriginIdentifier>ivo://vamdc/basecol/vamdc-tap_12.07</OriginIdentifier>
  </Origin>
\end{lstlisting}
The first line indicates that the resource producing the file is a VAMDC node and the second line indicates when the extraction has been performed (this time also corresponds to the creation of the file). On lines 3 to line 12 we find the information about the versioning of the contained data: it is specified that 
\begin{itemize}
\item the species referenced by the identifiers {\it XBAS2, XBAS52} (defined respectively at lines 56 and 89 of \ref{besecolFile});
\item the states referenced by the identifiers {\it SBASET54-1, SBASET52-1, SBASET52-2} (defined respectively at lines 58, 119 and 134  of \ref{besecolFile});
\item the processes referenced by the identifier {\it PBASC50t2T1c1C1} (defined at line 154 of \ref{besecolFile})
\item the sources referenced by the identifiers {\it BBAS0, BBAS849} (defined respectively at line 194 and 207 of \ref{besecolFile}.
\end{itemize}
belong to the version {\it VER001} of the database. This version was made available on the date {\tt 2015-09-01T08:10:12+01:00}.\\

Starting from line 13 we have the information about the database that produced the data: the link to its homepage (line 13), its Name (line 14), the query used for generating extracting the data and generating the file (lines 15 to 16), the unique identifier of the resource as it appears into the VAMDC infrastructure registry (line 17).\\

We recall that the two bibliographic references  {\it BBAS0} and {\it BBAS849}  included into this {\it VER001} version (cf. the last bullet of the previous item list) are respectively defined at line 194 of \ref{besecolFile}:
\begin{lstlisting}[style=listXML]
<Source sourceID="BBAS0">
      <Category>database</Category>
      <SourceName>BASECOL database</SourceName>
      <Year>2015</Year>
      <Authors>
        <Author>
          <Name>M.-L. Dubernet</Name>
        </Author>
      </Authors>
      <UniformResourceIdentifier>http://basecol.obspm.fr</UniformResourceIdentifier>
      <ProductionDate>2015-12-03</ProductionDate>
      <Comments>select * where ((target.MoleculeStoichiometricFormula = 'CO')) AND ((collider.AtomSymbol = 'he'))</Comments>
    </Source>  
\end{lstlisting}
and line 207 of \ref{besecolFile}:
\begin{lstlisting}[style=listXML]
<Source sourceID="BBAS849">
       <Category>journal</Category>
       <SourceName>apj</SourceName>
       <Year>2002</Year>
       <Authors>
         <Author>
           <Name>N. Balakrishnan</Name>
         </Author>
         <Author>
           <Name>A. Dalgarno</Name>
         </Author>
         <Author>
           <Name>C. Cecchi-Pestellini</Name>
         </Author>
         <Author>
           <Name>E. Bodo</Name>
         </Author>
       </Authors>
       <Title>Rotational and Vibrational Excitation of CO Molecules by Collisions with $^{4}$He Atoms</Title>
       <Volume>571</Volume>
       <PageBegin>1015</PageBegin>
       <PageEnd>1020</PageEnd>
       <UniformResourceIdentifier>
         http://adsabs.harvard.edu/cgi-bin/nph-bib_query?
         bibcode=2002JChPh.116.4517K&db_key=PHY
      </UniformResourceIdentifier>
     </Source>
\end{lstlisting}
This last reference is \cite{Pestellini}

\subsection{Example of an extraction from the CDMS node}
Let us now try to extract from the VAMDC infrastructure some spectroscopic data for the carbon monoxyde in a range of wavelength between $2.6006 \,10^7$ \AA  \, and 
$2.6008\, 10^7$ \AA. The associates query (typically generated by the portal from the Graphical User Web Interface and passed to all the VAMDC nodes) is \\ \\
{\tt select * where \\
(RadTransWavelength >= 2.6006E7 AND \\
RadTransWavelength <= 2.6008E7) AND \\
((MoleculeStoichiometricFormula = 'CO'))\\
}

The file reported in its entirety in \ref{cdmsFile} is the output that the CDMS node \cite{cdms1} generates when answering  the previous query (with the reservations made at the beginning of paragraph \ref{xsamsExamples}). Let us browse the content of this file and focus on the {\it Origin} element:
\begin{lstlisting}[style=listXML]
  <Origin xsi:type="VamdcNodeOriginType">  
    <Timestamp>2015-12-03T15:50:21+01:00</Timestamp>
    <Version versionID="VERCDMS1" global="false" timestamp="2000-10-01T12:00:00+01:00">
      <SpeciesRef>XCDMS-83</SpeciesRef>
      <StateRef>SCDMS-83-1</StateRef>
      <StateRef>SCDMS-83-2</StateRef>
      <StateRef>SCDMS-origin-83</StateRef>
      <ProcessRef>PCDMS-R15140649</ProcessRef>
      <SourceRef>BCDMS0</SourceRef>
      <SourceRef>BCDMS-1921</SourceRef>
       <SourceRef>BCDMS-1681</SourceRef>	   
    </Version>   
    <HomepageUrl>http://cdms.ph1.uni-koeln.de/</HomepageUrl>
    <Name>CDMS database</Name>
    <Query>select * where (RadTransWavelength &gt;= 2.6006E7 AND RadTransWavelength &lt;= 2.6008E7) AND ((MoleculeStoichiometricFormula = 'CO'))</Query>  
    <OriginIdentifier>ivo://vamdc/cdms/vamdc-tap_12.07</OriginIdentifier>
  </Origin>
\end{lstlisting}
As in the example from BASECOL, the first line indicates that the resource producing the file is a VAMDC node and the second line indicates when the file was generated. From line 3 to 12 we find the information about the versioning of the contained data: it is specified that 
\begin{itemize}
\item the species referenced by the identifier {\it XCDMS-83} (defined at line 24 of \ref{cdmsFile});
\item the states referenced by the identifiers {\it SCDMS-83-1, SCDMS-83-2, SCDMS-origin-83} (respectively defined at lines 67, 83, 51 of \ref{cdmsFile});
\item the process referenced by the identifier {\it PCDMS-R15140649} (defined at line 105 of \ref{cdmsFile})
\item the sources referenced by the identifier {\it BCDMS0, BCDMS-1921, BCDMS-1681} (defined respectively at line 131, 150 and 170 of \ref{cdmsFile})
\end{itemize}
belong to the version {\it VER1} of the database. This version was made available on the date {\tt 2000-10-01T12:00:00+01:00}.\\
Starting from line 13 we have the information about the database that produced the data: the link to its homepage (line 13), its name (line 14), the query used for extracting the data and generating the file (line 13), the unique identifier of the resource, as it appears into the VAMDC infrastructure registry (line 15).\\

We recall that the three bibliographic references {\it BCDMS0, BCDMS-1921} and {\it BCDMS-1681} (cf. the last bullet of the previous item list) are respectively defined at line 
131 of \ref{cdmsFile}
\begin{lstlisting}[style=listXML]
 <Source sourceID="BCDMS0">
    <Comments>This Source is a self-reference.
    It represents the database and the query that produced the xml document.
    The sourceID contains a timestamp.
    The full URL is given in the tag UniformResourceIdentifier but you need
    to unescape ampersands and angle brackets to re-use it.
    Query was:  select * where ((target.MoleculeStoichiometricFormula = 'CO')) AND ((collider.AtomSymbol = 'he'))</Comments>
         <Year>2015</Year>
         <Category>database</Category>
         <UniformResourceIdentifier>http://cdms.ph1.uni-koeln.de/cdms/tap/sync?
         LANG=VSS2&amp;
         REQUEST=doQuery&amp;FORMAT=XSAMS&amp;QUERY= select * where ((target.MoleculeStoichiometricFormula = 'CO')) AND ((collider.AtomSymbol = 'he'))</			UniformResourceIdentifier>
         <ProductionDate>2015-12-03</ProductionDate>
         <Authors>
            <Author>
               <Name>N.N.</Name>
            </Author>
         </Authors>
      </Source>
\end{lstlisting}
at line 150 of \ref{cdmsFile}
\begin{lstlisting}[style=listXML]
 <Source sourceID="BCDMS-1921">
         <Authors>
            <Author>
               <Name>Müller, H. S. P.</Name>
            </Author>
            <Author>
               <Name>Endres, C. P.</Name>
            </Author>
            <Author>
               <Name>Schlemmer, S.</Name>
            </Author>
            <Author>
               <Name>Stutzki, J.</Name>
            </Author>
         </Authors>
         <Title />
         <Category>database</Category>
         <Year>2012</Year>
         <SourceName>CDMS database</SourceName>
      </Source>
 \end{lstlisting}
and line 170 of \ref{cdmsFile}
\begin{lstlisting}[style=listXML]
 <Source sourceID="BCDMS-1681">
         <Authors>
            <Author>
               <Name>Winnewisser, G.</Name>
            </Author>
            <Author>
               <Name>Belov, S. P.</Name>
            </Author>
            <Author>
               <Name>Klaus, T.</Name>
            </Author>
            <Author>
               <Name>Schieder, R.</Name>
            </Author>
         </Authors>
         <Title />
         <Category>journal</Category>
         <Year>1997</Year>
         <SourceName>J. Mol. Spectrosc.</SourceName>
         <Volume>184</Volume>
         <PageBegin>468</PageBegin>
         <DigitalObjectIdentifier>
         	10.1006/jmsp.1997.7341
	</DigitalObjectIdentifier>
      </Source>
\end{lstlisting}
This last reference is \cite{1997JMoSp}.

\subsection{Documenting data processing workflow: merging of two files with SpectCol}
We used the SpectCol tool \cite{spectcol1},\cite{spectcol2}\footnote{http://www.vamdc.org/software/spectcol} for cross-matching the data from the two files described into the previous subparagraphs.  The SpectCol tool allows to create combined spectroscopic and collisional data files where the collisional
rate coefficients come from a collisional database (BASECOL in this example) and the spectroscopic data (energy levels and 
Einstein coefficients) come from a spectroscopic database (CDMS in this example). This process requires to match the
spectroscopic energy levels of CDMS with those of BASECOL on the basis of identical quantum numbers.
We asked Spectcol to merge the two files by identifying the states of the Carbon Monoxyde characterised by the same quantum number values $J$: the state referenced by the identifier {\it SBASET52-1} ($J=0$, cf. line 117 of \ref{basecolOutput}) has been identified and assimilated with {\it CDMS-83-1} (cf. line 64 of \ref{cdmsFile}) while the state referenced by the identifier {\it SBASET52-2} ($J=1$, cf. line 132 of \ref{basecolOutput}) has been identified and assimilated with {\it CDMS-83-1} (cf. line 80 of \ref{cdmsFile}). \\
The resulting merged file is reported in \ref{merged}. Let us browse the content of this file and focus on the root {\it Origin} element:
\begin{lstlisting}[style=listXML]
  <Origin xsi:type="OtherOriginType">  
    <Timestamp>2015-12-07T15:50:21+01:00</Timestamp>
    <Version versionID="VERMER1" global="false" timestamp="2000-10-01T12:00:00+01:00">
      <ProcessRef>PBASC50t2T1c1C1</ProcessRef>
      <SourceRef>BBAS0</SourceRef>
      <SourceRef>BBAS849</SourceRef>
      <SourceRef>BCDMS0</SourceRef>
      <SourceRef>BCDMS-1921</SourceRef>
      <SourceRef>BCDMS-1681</SourceRef>
    </Version>   
    <HomepageUrl>
    	http://www.vamdc.org/activities/research/software/spectcol/
    </HomepageUrl>
    <Name>Spectcol</Name>
    <Comments>Merging between collisional data from Basecol and spectroscopic data from CDMS by cross matching state with same J values in CO</Comments>
    
    <Origin xsi:type="VamdcNodeOriginType">  
      <Timestamp>2015-12-03T15:50:21+01:00</Timestamp>
      <Version versionID="VERCDMS1" global="false" timestamp="2000-10-01T12:00:00+01:00">
        <SpeciesRef>XCDMS-83</SpeciesRef>
        <StateRef>SCDMS-83-1</StateRef>
        <StateRef>SCDMS-83-2</StateRef>
        <StateRef>SCDMS-origin-83</StateRef>
        <ProcessRef>PCDMS-R15140649</ProcessRef>
      </Version>   
      <HomepageUrl>http://cdms.ph1.uni-koeln.de/</HomepageUrl>
      <Name>CDMS database</Name>
      <Comments>Spectroscopic data extracted from CDMS VAMDC node.</Comments>
      <Query>select * where (RadTransWavelength &gt;= 2.6006E7 AND RadTransWavelength &lt;= 2.6008E7) AND ((MoleculeStoichiometricFormula = 'CO'))</Query>  
      <OriginIdentifier>ivo://vamdc/cdms/vamdc-tap_12.07</OriginIdentifier>      
    </Origin>
    
    <Origin xsi:type="VamdcNodeOriginType">
      <Timestamp>2015-12-03T14:40:21+01:00</Timestamp>
      <Version versionID="VER001" global="false" timestamp="2015-09-01T08:10:12+01:00">
        <SpeciesRef>XBAS2</SpeciesRef>
        <StateRef>SBASET54-1</StateRef>        
      </Version>
      <HomepageUrl>http://basecol.vamdc.org</HomepageUrl>
      <Name>Basecol</Name>
      <Comments>Collisional data extracted from Basecol VAMDC node.</Comments>
      <Query>select * where ((target.MoleculeStoichiometricFormula = 'CO')) AND ((collider.AtomSymbol = 'he'))</Query>    
      <OriginIdentifier>ivo://vamdc/basecol/vamdc-tap_12.07</OriginIdentifier>
    </Origin>
    
  </Origin>
\end{lstlisting}
We see (line 1) that the file was not produced by a VAMDC node, but by the Spectcol tool (line 11 to 15). The current file has been produced on {\tt 2015-12-07T15:50:21+01:00} and results from the merging of data coming from BASECOL and from CDMS, where the states of $CO$ characterised by the same values of the quantum number $J$ are crossmatched (line 15). The data coming from BASECOL and CDMS are detailed  into the two contained  {\it Origin} sub-elements:
\begin{itemize}
\item the first one is defined from line 17 to line 31 and holds the data coming from the CDMS node. This includes the species referenced by the identifier {\it XCDMS-83}, the states referenced by the identifier {\it SCDMS-83-1, SCDMS-83-2, SCDMS-origin-83} (respectively defined at line 137, 151, 165 of \ref{merged}), the process referenced by the identifier {\it PCDMS-R15140649} (defined at line 186 of \ref{merged}). The extraction of those data was performed on {\tt 2015-12-03T15:50:21+01:00} from the CDMS Node  by submitting the query \\
{\tt select * where \\
(RadTransWavelength >= 2.6006E7 AND \\
RadTransWavelength <= 2.6008E7) AND \\
((MoleculeStoichiometricFormula = 'CO'))\\
}
The extracted data correspond to the version {\it VERCDMS1} that was available to the community since {\tt 2000-10-01T12:00:00+01:00}
\item the second one is defined from line 31 to line 42 and holds the data coming from the BASECOL node. This includes the species referenced by the identifier {\it XBAS2} (defined at line 75 of \ref{merged}) and the state referenced by the identifier {\it SBASET54-1} (defined at line 77 of \ref{merged}). The extraction of those data was performed on {\tt 2015-12-03T14:40:21+01:00} from the BASECOL node by submitting the query \\
{\tt select * where \\((target.MoleculeStoichiometricFormula = 'CO')) AND \\((collider.AtomSymbol = 'he'))}\\
The extracted data correspond to the version {\it VER001} of the database. This was available to the community since {\tt 2015-09-01T08:10:12+01:00}.
\end{itemize}
Let us focus again on the root {\it Origin} element:
\begin{itemize}
\item As we can see from line 5 to 9, all the sources references from the merged files are now contained into the {\it Version} object of the root {\it Origin}. Indeed during the merging procedure, the sources from both the BASECOL and the CDMS files are compiled together into the root element since the data coming from all the sources are used. Moreover for data citation,  it is very convenient to store at a same location all the references linked to the current file.
\item The process referenced by {\it PBASC50t2T1c1C1} belongs to the {\it Version} field of the root element (line 4). If we compare with \ref{besecolFile}, line 154, we discover that this is a collisional process coming from BASECOL. It is important to understand why it is not contained into the second sub-origin element which, as we previously shown, contains the data coming from BASECOL: in its original extraction from BASECOL (cf. \ref{basecolOutput}, line 154) this process is a collision between $CO$ (identified with the label {\it XBAS52}) and $He$ (identified with the label {\it XBAS2}). During the collision, the $CO$ pass from the state referenced by {\it SBASET52-2} to the state referenced by {\it SBASET52-1}. All these identifiers and references have consistency in the BASECOL perimeter. Let us look at the same collisional process as it is described on the merged file (cf. line 212 of \ref{merged}): the collision is between $CO$ (now identified with the label {\it XCDMS-83}, since during the crossmatching procedure we have replaced the $CO$ from BASECOL with the one coming from CDMS) and $He$. During the collision $CO$ pass from the state referenced by {\it SCDMS-83-2} to the state referenced by {\it SCDMS-83-1} (indeed we remember that during the crossmatching we have identified the state {\it SBASET52-2} from BASECOL with the state {\it SCDMS-83-2} from CDMS and the states {\it SBASET52-1} from BASECOL with the state {\it SCDMS-83-1} from CDMS). Due to these changes, the collisional process refer now to data coming both from BASECOL and CDMS and the data is no more the original one freshly extracted from BASECOL (as it was the case in \ref{basecolOutput}, line 154). As a direct consequence BASECOL is no more the  source origin of this data. This combined collision is created by SpectCol and it is natural that it appears into the root {\it Origin} element which holds the data produced by SpectCol during the crossmatching procedure. 
\end{itemize}

\section{Practical use of the new XSAMS features}\label{queryStore}
In section \ref{xsamsEvol} we introduced new changes and additions to the XSAMS format which allows one to identify unequivocally through time and location the datasets extracted from VAMDC: all the relevant metadata (the extraction query, the date when the query was submitted to the infrastructure, the responding node, the version of the responding node, etc...) are automatically embedded into the output file. But from the practical point of view of the final user, how to use this information ? Our aim is to simplify the ``data citation'' landscape and not to complicate the actual one described in section \ref{existingSolution}. In this paragraph we are going to sketch the future VAMDC services that will allow a simple and fluid use of these new features: \\
When a user  queries data from the VAMDC e-infrastructure by submitting a given query, he/she will receive an output file formatted as described in section \ref{xsamsEvol}. This file will be associated with a unique digital identifier. This identifier (as it is the case for the DOI), will be resolvable and the landing page\footnote{The landing page will be human readable, but a machine actionable version may exist for automatic treatements} will summarize the information contained in the {\it Origin} field of the output file, the data-file itself (via query re-execution, if the node supports this feature)\footnote{The evolved XSAMS standard described in section \ref{xsamsEvol}  allows a data provider to univocally tag version of their data. Even if we strongly suggest that the VAMDC nodes  give access (through the VAMC infrastructure) to all the subsequent versions of their data, we cannot enforce the data policy on the federated nodes. A given node may delete an old version when a new one is available. In that case the metadata policy will always allow the unambiguous identification of the extracted data, but data extraction re-execution will not be possible}, the bibtex references of all the papers used for compiling the output file. 
This sketched service is a part of  a {\it Query Store} as it is understood in the Data Citation RDA context and described in \cite{rdawgdc-recdocument}.\\

This query store may be used in collaboration with editors. We imagine the following data-citation workflow: when a scientist wishes to cite a dataset extracted from VAMDC, he/she has just to put into the references the {\it unique identifier} returned by the VAMDC infrastructure during the extraction procedure. Since this unique identifier is resolvable, the editors may resolve it and, by accessing the bibtex information of the landing page, give credits to all the listed authors. This automatic credit delegation method may also work with current systems (e.g. the SAO/NASA Astrophysics Data System\footnote{http://adswww.harvard.edu/}).

\section{Concluding remarks}
Through this work we have discussed an evolution of the VAMDC standard output format which provides a mechanisms allowing to identify through time and location, in a sustainable and unique way, data sets extracted by users from the VAMDC e-infrastructure. With our proposed solution the information about the origin and version of data are embedded into (and indissoluble from) the file containing the atomic and molecular data. The same mechanisms define an unambiguous method for tracing and documenting the data processing workflows. \\ 

The methods and technical solutions introduced in this paper are designed for scientist wishing to efficiently cite the datasets and the data-producers of the Atomic and Molecular data extracted from the VAMDC infrastructure: \\
let us consider an XSAMS data file containing $N$ references ($N >>1$) to the sources used for the data file compilation.\\ 
Before the introduction of the solution discussed into this paper, a scientist wishing to publish a work based on that data file would need to include the $N$ publications in order to properly cite the data.\\
With the introduction of the proposed solution, the scientist will only cite the Digital Unique Identifier (DUI) returned by the VAMDC infrastructure during the extraction procedure. This DUI is resolvable and the landing page will contain the original query used for the generation of the data file and the bibliographic references to the $N$ publications.

\section{ Acknowledgements}
Support for VAMDC has been provided through the VAMDC and the
  SUP@VAMDC projects funded under the ``Combination of Collaborative
  Projects and Coordination and Support Actions'' Funding Scheme of
  The Seventh Framework Program. Call topic: INFRA-2008-1.2.2 and
  INFRA-2012 Scientific Data Infrastructure. Grant Agreement numbers:
  239108 and 313284. We acknowledge support from Paris
  Astronomical Data Center. {\bf We thank Thomas Marquart, Guy Rixon and Mickail Doronin from the VAMDC Consortium for
  fruitful discussions}.

\appendix

\section{Sample output from BASECOL}\label{besecolFile}
\label{basecolOutput}
\begin{lstlisting}[style=listXML]
<?xml version="1.0" encoding="UTF-8" standalone="yes"?>
<XSAMSData xmlns="http://vamdc.org/xml/xsams/1.0" 
  xmlns:cml="http://www.xml-cml.org/schema" 
  xmlns:asymcs="http://vamdc.org/xml/xsams/1.0/cases/asymcs" 
  xmlns:asymos="http://vamdc.org/xml/xsams/1.0/cases/asymos" 
  xmlns:dcs="http://vamdc.org/xml/xsams/1.0/cases/dcs" 
  xmlns:gen="http://vamdc.org/xml/xsams/1.0/cases/gen" 
  xmlns:hunda="http://vamdc.org/xml/xsams/1.0/cases/hunda" 
  xmlns:hundb="http://vamdc.org/xml/xsams/1.0/cases/hundb" 
  xmlns:lpcs="http://vamdc.org/xml/xsams/1.0/cases/lpcs" 
  xmlns:lpos="http://vamdc.org/xml/xsams/1.0/cases/lpos" 
  xmlns:ltcs="http://vamdc.org/xml/xsams/1.0/cases/ltcs" 
  xmlns:ltos="http://vamdc.org/xml/xsams/1.0/cases/ltos" 
  xmlns:nltcs="http://vamdc.org/xml/xsams/1.0/cases/nltcs" 
  xmlns:nltos="http://vamdc.org/xml/xsams/1.0/cases/nltos" 
  xmlns:sphcs="http://vamdc.org/xml/xsams/1.0/cases/sphcs" 
  xmlns:sphos="http://vamdc.org/xml/xsams/1.0/cases/sphos" 
  xmlns:stcs="http://vamdc.org/xml/xsams/1.0/cases/stcs"
  xmlns:xsi="http://www.w3.org/2001/XMLSchema-instance"
  xsi:schemaLocation="http://vamdc.org/xml/xsams/1.0 
	https://raw.githubusercontent.com/nicolasmoreau/
	VAMDC-XSAMS/abstract_origin/xsams.xsd">
  
   <Origin xsi:type="VamdcNodeOriginType">
    <Timestamp>2015-12-03T14:40:21+01:00</Timestamp>
    <Version versionID="VER001" global="false" timestamp="2015-09-01T08:10:12+01:00">
      <SpeciesRef>XBAS2</SpeciesRef>
      <SpeciesRef>XBAS52</SpeciesRef>
      <StateRef>SBASET54-1</StateRef>
      <StateRef>SBASET52-1</StateRef>
      <StateRef>SBASET52-2</StateRef>
      <ProcessRef>PBASC50t2T1c1C1</ProcessRef>
      <SourceRef>BBAS0</SourceRef>
      <SourceRef>BBAS849</SourceRef>
    </Version>
    <HomepageUrl>http://basecol.vamdc.org</HomepageUrl>
    <Name>Basecol</Name>
    <Query>select * where ((target.MoleculeStoichiometricFormula = 'CO')) AND 
    ((collider.AtomSymbol = 'he'))</Query>    
    <OriginIdentifier>ivo://vamdc/basecol/vamdc-tap_12.07</OriginIdentifier>
  </Origin>
  <Species>
    <Atoms>
      <Atom>
        <ChemicalElement>
          <NuclearCharge>2</NuclearCharge>
          <ElementSymbol>He</ElementSymbol>
        </ChemicalElement>
        <Isotope>
          <IsotopeParameters>
            <MassNumber>4</MassNumber>
            <Mass>
              <Value units="amu">4.0026</Value>
            </Mass>
          </IsotopeParameters>
          <Ion speciesID="XBAS2">
            <IonCharge>0</IonCharge>
            <AtomicState stateID="SBASET54-1">
              <Comments>Energy level of He (no structure)</Comments>
              <SourceRef>BBAS0</SourceRef>
              <AtomicNumericalData>
                <StateEnergy>
                  <Value units="1/cm">0.0</Value>
                </StateEnergy>
              </AtomicNumericalData>
              <AtomicQuantumNumbers>
                <TotalAngularMomentum>0.0</TotalAngularMomentum>
              </AtomicQuantumNumbers>
              <AtomicComposition>
                <Component>
                  <Term>
                    <LS>
                      <L>
                        <Value>0</Value>
                        <Symbol>L</Symbol>
                      </L>
                      <S>0.0</S>
                    </LS>
                  </Term>
                </Component>
              </AtomicComposition>
            </AtomicState>
            <InChIKey>SWQJXJOGLNCZEY-UHFFFAOYSA-N</InChIKey>
          </Ion>
        </Isotope>
      </Atom>
    </Atoms>
    <Molecules>     
      <Molecule speciesID="XBAS52">
        <MolecularChemicalSpecies>
          <OrdinaryStructuralFormula>
            <Value>CO</Value>
          </OrdinaryStructuralFormula>
          <StoichiometricFormula>CO</StoichiometricFormula>
          <ChemicalName>
            <Value>CO</Value>
          </ChemicalName>
          <InChI>InChI=1S/CO/c1-2</InChI>
          <InChIKey>UGFAIRIUMAVXCW-UHFFFAOYSA-N</InChIKey>
          <VAMDCSpeciesID>UGFAIRIUMAVXCW-UHFFFAOYSA-N</VAMDCSpeciesID>
          <MoleculeStructure>
            <cml:atomArray>
              <cml:atom hydrogenCount="0" isotopeNumber="16" count="1.0" 
              formalCharge="1" elementType="O" id="RBAS43N357"/>
              <cml:atom hydrogenCount="0" isotopeNumber="12" count="1.0" 
              formalCharge="-1" elementType="C" id="RBAS43N358"/>
            </cml:atomArray>
            <cml:bondArray>
              <cml:bond order="T" atomRefs2="RBAS43N357 RBAS43N358"/>
            </cml:bondArray>
          </MoleculeStructure>
          <StableMolecularProperties>
            <MolecularWeight>
              <Value units="amu">27.99</Value>
            </MolecularWeight>
          </StableMolecularProperties>
          <Comment>Theoretical rotational energy levels of CO (Cecchi-Pestellini, 2002)</Comment>
        </MolecularChemicalSpecies>
        <MolecularState stateID="SBASET52-1">
          <SourceRef>BBAS0</SourceRef>
          <Description>Theoretical rotational energy levels of CO (Cecchi-Pestellini, 2002)</Description>
          <MolecularStateCharacterisation>
            <StateEnergy energyOrigin="SBASET52-1">
              <Value units="1/cm">0.0</Value>
            </StateEnergy>
          </MolecularStateCharacterisation>
          <Case xmlns:xsi="http://www.w3.org/2001/XMLSchema-instance" xsi:type="dcs:Case" caseID="dcs">
            <dcs:QNs>
              <dcs:ElecStateLabel>X</dcs:ElecStateLabel>
              <dcs:J>0</dcs:J>
            </dcs:QNs>
          </Case>
        </MolecularState>
        <MolecularState stateID="SBASET52-2">
          <SourceRef>BBAS0</SourceRef>
          <Description>Theoretical rotational energy levels of CO (Cecchi-Pestellini, 2002)</Description>
          <MolecularStateCharacterisation>
            <StateEnergy energyOrigin="SBASET52-1">
              <Value units="1/cm">3.85</Value>
            </StateEnergy>
          </MolecularStateCharacterisation>
          <Case xmlns:xsi="http://www.w3.org/2001/XMLSchema-instance" xsi:type="dcs:Case" caseID="dcs">
            <dcs:QNs>
              <dcs:ElecStateLabel>X</dcs:ElecStateLabel>
              <dcs:J>1</dcs:J>
            </dcs:QNs>
          </Case>
        </MolecularState>       
      </Molecule>
    </Molecules>
  </Species>
  <Processes>
    <Collisions>
      <CollisionalTransition id="PBASC50t2T1c1C1">
        <Comments> Rotational de-excitation of CO (v=0) by He (Cecchi-Pestellini &amp; al, 2002)</Comments>
        <SourceRef>BBAS0</SourceRef>
        <SourceRef>BBAS849</SourceRef>
        <ProcessClass>
          <Code>inel</Code>
        </ProcessClass>
        <Reactant>
          <SpeciesRef>XBAS52</SpeciesRef>
          <StateRef>SBASET52-2</StateRef>
        </Reactant>
        <Reactant>
          <SpeciesRef>XBAS2</SpeciesRef>
          <StateRef>SBASET54-1</StateRef>
        </Reactant>
        <Product>
          <SpeciesRef>XBAS52</SpeciesRef>
          <StateRef>SBASET52-1</StateRef>
        </Product>
        <Product>
          <SpeciesRef>XBAS2</SpeciesRef>
          <StateRef>SBASET54-1</StateRef>
        </Product>
        <DataSets>
          <DataSet dataDescription="rateCoefficient">
            <TabulatedData>
              <Comments>Rate coefficients</Comments>
              <X units="K">
                <DataList>5.0 10.0 20.0 40.0 60.0 80.0 100.0 200.0 300.0 500.0</DataList>
              </X>
              <Y units="cm3/s">
                <DataList>3.4E-11 3.2E-11 3.0E-11 2.8E-11 2.7E-11 2.6E-11 2.6E-11 2.5E-11 2.5E-11 2.6E-11</DataList>
              </Y>
            </TabulatedData>
          </DataSet>
        </DataSets>
      </CollisionalTransition>   
    </Collisions>
  </Processes>
  <Sources>
    <Source sourceID="BBAS0">
      <Category>database</Category>
      <SourceName>BASECOL database</SourceName>
      <Year>2015</Year>
      <Authors>
        <Author>
          <Name>M.-L. Dubernet</Name>
        </Author>
      </Authors>
      <UniformResourceIdentifier>http://basecol.obspm.fr</UniformResourceIdentifier>
      <ProductionDate>2015-12-03</ProductionDate>
      <Comments>select * where ((target.MoleculeStoichiometricFormula = 'CO')) AND ((collider.AtomSymbol = 'he'))</Comments>
    </Source>  
    <Source sourceID="BBAS849">
       <Category>journal</Category>
       <SourceName>apj</SourceName>
       <Year>2002</Year>
       <Authors>
         <Author>
           <Name>N. Balakrishnan</Name>
         </Author>
         <Author>
           <Name>A. Dalgarno</Name>
         </Author>
         <Author>
           <Name>C. Cecchi-Pestellini</Name>
         </Author>
         <Author>
           <Name>E. Bodo</Name>
         </Author>
       </Authors>
       <Title>Rotational and Vibrational Excitation of CO Molecules by Collisions with $^{4}$He Atoms</Title>
       <Volume>571</Volume>
       <PageBegin>1015</PageBegin>
       <PageEnd>1020</PageEnd>
       <UniformResourceIdentifier>
    http://adsabs.harvard.edu/cgi-bin/nph-bib_query?
         bibcode=2002JChPh.116.4517K&db_key=PHY
       </UniformResourceIdentifier>
     </Source>
  </Sources>
</XSAMSData>
\end{lstlisting}

\section{Sample output from CDMS}\label{cdmsFile}
\begin{lstlisting}[style=listXML]
<?xml version="1.0" encoding="UTF-8"?>
<XSAMSData xmlns="http://vamdc.org/xml/xsams/1.0" xmlns:xsi="http://www.w3.org/2001/XMLSchema-instance" 
xmlns:cml="http://www.xml-cml.org/schema" 
xsi:schemaLocation="http://vamdc.org/xml/xsams/1.0 https://raw.githubusercontent.com/nicolasmoreau/VAMDC-XSAMS/abstract_origin/xsams.xsd">
  <Origin xsi:type="VamdcNodeOriginType">  
    <Timestamp>2015-12-03T15:50:21+01:00</Timestamp>
    <Version versionID="VERCDMS1" global="false" timestamp="2000-10-01T12:00:00+01:00">
      <SpeciesRef>XCDMS-83</SpeciesRef>
      <StateRef>SCDMS-83-1</StateRef>
      <StateRef>SCDMS-83-2</StateRef>
      <StateRef>SCDMS-origin-83</StateRef>
      <ProcessRef>PCDMS-R15140649</ProcessRef>
      <SourceRef>BCDMS0</SourceRef>
      <SourceRef>BCDMS-1921</SourceRef>
      <SourceRef>BCDMS-1681</SourceRef>
    </Version>   
    <HomepageUrl>http://cdms.ph1.uni-koeln.de/</HomepageUrl>
    <Name>CDMS database</Name>
    <Query>select * where (RadTransWavelength &gt;= 2.6006E7 AND RadTransWavelength &lt;= 2.6008E7) AND ((MoleculeStoichiometricFormula = 'CO'))</Query>  
    <OriginIdentifier>ivo://vamdc/cdms/vamdc-tap_12.07</OriginIdentifier>
  </Origin>
  <Species>
    <Molecules>
      <Molecule speciesID="XCDMS-83">
        <MolecularChemicalSpecies>
          <OrdinaryStructuralFormula>
            <Value>CO</Value>
          </OrdinaryStructuralFormula>
          <StoichiometricFormula>CO</StoichiometricFormula>
          <ChemicalName>
            <Value>Carbon Monoxide</Value>
          </ChemicalName>
          <InChI>1S/CO/c1-2</InChI>
          <InChIKey>UGFAIRIUMAVXCW-UHFFFAOYSA-N</InChIKey>
          <VAMDCSpeciesID>UGFAIRIUMAVXCW-UHFFFAOYSA-N</VAMDCSpeciesID>
          <PartitionFunction>
            <T units="K">
              <DataList>1.072 1.148 1.23 1.318 1.413 1.514 1.622 1.738 1.862 1.995 2.138 2.291 2.455 2.63 2.725 2.818 3.02 3.236 3.467 3.715 3.981 4.266 4.571 4.898 5.0 5.248 5.623 6.026 6.457 6.918 7.413 7.943 8.511 9.12 9.375 9.772 10.471 11.22 12.023 12.882 13.804 14.791 15.849 16.982 18.197 18.75 19.498 20.893 22.387 23.988 25.704 27.542 29.512 31.623 33.884 36.308 37.5 38.905 41.687 44.668 47.863 51.286 54.954 58.884 63.096 67.608 72.444 75.0 77.625 83.176 89.125 95.499 102.329 109.648 117.49 125.893 134.896 144.544 150.0 154.882 165.959 177.828 190.546 204.174 218.776 225.0 234.423 251.189 269.153 288.403 300.0 309.03 331.131 354.813 380.189 407.38 436.516 467.735 500.0 501.187 537.032 575.44 616.595 660.693 707.946 758.578 812.831 870.964 933.254 1000.0</DataList>
            </T>
            <Q>
              <DataList>1.01721581835 1.02422837677 1.03341102522 1.04512141245 1.05985020123 1.07774884294 1.09923417933 1.12473229226 1.15441650589 1.18864057708 1.22774305695 1.27174864212 1.32092124733 1.37520246158 1.40530654907 1.43514349042 1.50100121324 1.57270208471 1.65050158216 1.73501239088 1.8265276783 1.92535019369 2.03179140659 2.14652192999 2.18241666936 2.26987115366 2.40252640741 2.54553752733 2.69889427601 2.8632973732 3.04016627563 3.22985295285 3.4334258678 3.65195638949 3.74352703447 3.88615863094 4.13746641418 4.40695607053 4.69606494313 5.00551115749 5.33781506844 5.69369660756 6.0753182363 6.48412210856 6.9226335574 7.12225433182 7.3922960322 7.89599749116 8.4355434648 9.0138233311 9.633727213 10.2977845092 11.0096090065 11.7724538098 12.5895724935 13.4656648032 13.8965040828 14.4043468795 15.4099582553 16.4875620501 17.6425836137 18.8800875183 20.2062236736 21.6271428763 23.1500815655 24.7815541073 26.5302455525 27.4545173427 28.40375754 30.4111396059 32.5625279126 34.8676989471 37.3378779761 39.9850161394 42.8214291179 45.8608826253 49.1175078254 52.6076105394 54.5813579159 56.3475015965 60.3549445917 64.6491571735 69.2508130864 74.1820459087 79.4660945815 81.7185083304 85.1287658147 91.1970158766 97.6997287418 104.669260558 108.868854329 112.139428854 120.146852376 128.732781329 137.942190312 147.825233889 158.438549717 169.84604324 181.685035016 182.121688241 195.35029188 209.62906021 225.070653126 241.803934912 259.976802963 279.755720344 301.328756805 324.90561697 350.71803967 379.021079637</DataList>
            </Q>
          </PartitionFunction>
          <StableMolecularProperties>
            <MolecularWeight>
              <Value units="unitless">28</Value>
            </MolecularWeight>
          </StableMolecularProperties>
          <Comment> 28503- v 1:CO; $v=0$</Comment>
        </MolecularChemicalSpecies>
        <MolecularState auxillary="true" stateID="SCDMS-origin-83">
          <MolecularStateCharacterisation>
            <StateEnergy energyOrigin="SCDMS-origin-83">
              <Value units="1/cm">0.0</Value>
            </StateEnergy>
            <TotalStatisticalWeight>1</TotalStatisticalWeight>
            <NuclearStatisticalWeight>1</NuclearStatisticalWeight>
          </MolecularStateCharacterisation>
          <Case xmlns:dcs="http://vamdc.org/xml/xsams/1.0/cases/dcs" xsi:type="dcs:Case" caseID="dcs" xsi:schemaLocation="http://vamdc.org/xml/xsams/1.0/cases/dcs ../../cases/dcs.xsd">
            <dcs:QNs>
              <dcs:ElecStateLabel>X</dcs:ElecStateLabel>
              <dcs:v>0</dcs:v>
              <dcs:J>0</dcs:J>
            </dcs:QNs>
          </Case>
        </MolecularState>
        <MolecularState stateID="SCDMS-83-1">
          <MolecularStateCharacterisation>
            <StateEnergy energyOrigin="SCDMS-origin-83">
              <Value units="1/cm">0.0</Value>
            </StateEnergy>
            <TotalStatisticalWeight>1</TotalStatisticalWeight>
            <NuclearStatisticalWeight>1</NuclearStatisticalWeight>
          </MolecularStateCharacterisation>
          <Case xmlns:dcs="http://vamdc.org/xml/xsams/1.0/cases/dcs" xsi:type="dcs:Case" caseID="dcs" xsi:schemaLocation="http://vamdc.org/xml/xsams/1.0/cases/dcs ../../cases/dcs.xsd">
            <dcs:QNs>
              <dcs:ElecStateLabel>X</dcs:ElecStateLabel>
              <dcs:v>0</dcs:v>
              <dcs:J>0</dcs:J>
            </dcs:QNs>
          </Case>
        </MolecularState>
        <MolecularState stateID="SCDMS-83-2">
          <MolecularStateCharacterisation>
            <StateEnergy energyOrigin="SCDMS-origin-83">
              <Value units="1/cm">3.845033</Value>
            </StateEnergy>
            <TotalStatisticalWeight>3</TotalStatisticalWeight>
            <NuclearStatisticalWeight>1</NuclearStatisticalWeight>
          </MolecularStateCharacterisation>
          <Case xmlns:dcs="http://vamdc.org/xml/xsams/1.0/cases/dcs" xsi:type="dcs:Case" caseID="dcs" xsi:schemaLocation="http://vamdc.org/xml/xsams/1.0/cases/dcs ../../cases/dcs.xsd">
            <dcs:QNs>
              <dcs:ElecStateLabel>X</dcs:ElecStateLabel>
              <dcs:v>0</dcs:v>
              <dcs:J>1</dcs:J>
            </dcs:QNs>
          </Case>
        </MolecularState>
      </Molecule>
    </Molecules>
  </Species>
  <Processes>
    <Radiative>
      <RadiativeTransition id="PCDMS-R15140649" process="excitation">
        <SourceRef>BCDMS0</SourceRef>
        <EnergyWavelength>
          <Frequency>
            <Value units="MHz">115271.2021</Value>
            <Accuracy>0.0001</Accuracy>
          </Frequency>
        </EnergyWavelength>
        <UpperStateRef>SCDMS-83-2</UpperStateRef>
        <LowerStateRef>SCDMS-83-1</LowerStateRef>
        <SpeciesRef>XCDMS-83</SpeciesRef>
        <Probability>
          <TransitionProbabilityA>
            <Value units="1/cm">7.20360334988e-08</Value>
          </TransitionProbabilityA>
          <IdealisedIntensity>
            <Value units="unitless">-5.0105</Value>
          </IdealisedIntensity>
          <Multipole>E2</Multipole>
        </Probability>
        <ProcessClass>
          <Code>rota</Code>
        </ProcessClass>
      </RadiativeTransition>
    </Radiative>
  </Processes>
  <Sources>
    <Source sourceID="BCDMS0">
    <Comments>This Source is a self-reference.
    It represents the database and the query that produced the xml document.
    The sourceID contains a timestamp.
    The full URL is given in the tag UniformResourceIdentifier but you need
    to unescape ampersands and angle brackets to re-use it.
    Query was:  select * where ((target.MoleculeStoichiometricFormula = 'CO')) AND ((collider.AtomSymbol = 'he'))</Comments>
         <Year>2015</Year>
         <Category>database</Category>
         <UniformResourceIdentifier>http://cdms.ph1.uni-koeln.de/cdms/tap/sync?
         LANG=VSS2&amp;
         REQUEST=doQuery&amp;FORMAT=XSAMS&amp;QUERY= select * where ((target.MoleculeStoichiometricFormula = 'CO')) AND ((collider.AtomSymbol = 'he'))</			UniformResourceIdentifier>
         <ProductionDate>2015-12-03</ProductionDate>
         <Authors>
            <Author>
               <Name>N.N.</Name>
            </Author>
         </Authors>
      </Source>
      <Source sourceID="BCDMS-1921">
         <Authors>
            <Author>
               <Name>Müller, H. S. P.</Name>
            </Author>
            <Author>
               <Name>Endres, C. P.</Name>
            </Author>
            <Author>
               <Name>Schlemmer, S.</Name>
            </Author>
            <Author>
               <Name>Stutzki, J.</Name>
            </Author>
         </Authors>
         <Title />
         <Category>database</Category>
         <Year>2012</Year>
         <SourceName>CDMS database</SourceName>
      </Source>
      <Source sourceID="BCDMS-1681">
         <Authors>
            <Author>
               <Name>Winnewisser, G.</Name>
            </Author>
            <Author>
               <Name>Belov, S. P.</Name>
            </Author>
            <Author>
               <Name>Klaus, T.</Name>
            </Author>
            <Author>
               <Name>Schieder, R.</Name>
            </Author>
         </Authors>
         <Title />
         <Category>journal</Category>
         <Year>1997</Year>
         <SourceName>J. Mol. Spectrosc.</SourceName>
         <Volume>184</Volume>
         <PageBegin>468</PageBegin>
         <DigitalObjectIdentifier>10.1006/jmsp.1997.7341</DigitalObjectIdentifier>
      </Source>
  </Sources>
</XSAMSData>
\end{lstlisting}

\section{Merged output from SpectCol}\label{merged}
\begin{lstlisting}[style=listXML]
<?xml version="1.0" encoding="UTF-8" standalone="yes"?>
<XSAMSData xmlns="http://vamdc.org/xml/xsams/1.0" 
  xmlns:cml="http://www.xml-cml.org/schema" 
  xmlns:asymcs="http://vamdc.org/xml/xsams/1.0/cases/asymcs" 
  xmlns:asymos="http://vamdc.org/xml/xsams/1.0/cases/asymos" 
  xmlns:dcs="http://vamdc.org/xml/xsams/1.0/cases/dcs" 
  xmlns:gen="http://vamdc.org/xml/xsams/1.0/cases/gen" 
  xmlns:hunda="http://vamdc.org/xml/xsams/1.0/cases/hunda" 
  xmlns:hundb="http://vamdc.org/xml/xsams/1.0/cases/hundb" 
  xmlns:lpcs="http://vamdc.org/xml/xsams/1.0/cases/lpcs" 
  xmlns:lpos="http://vamdc.org/xml/xsams/1.0/cases/lpos"
  xmlns:ltcs="http://vamdc.org/xml/xsams/1.0/cases/ltcs" 
  xmlns:ltos="http://vamdc.org/xml/xsams/1.0/cases/ltos" 
  xmlns:nltcs="http://vamdc.org/xml/xsams/1.0/cases/nltcs" 
  xmlns:nltos="http://vamdc.org/xml/xsams/1.0/cases/nltos" 
  xmlns:sphcs="http://vamdc.org/xml/xsams/1.0/cases/sphcs" 
  xmlns:sphos="http://vamdc.org/xml/xsams/1.0/cases/sphos" 
  xmlns:stcs="http://vamdc.org/xml/xsams/1.0/cases/stcs"
  xmlns:xsi="http://www.w3.org/2001/XMLSchema-instance" 
  xsi:schemaLocation="http://vamdc.org/xml/xsams/1.0 https://raw.githubusercontent.com/nicolasmoreau/VAMDC-XSAMS/abstract_origin/xsams.xsd"
  >
  <Origin xsi:type="OtherOriginType">  
    <Timestamp>2015-12-07T15:50:21+01:00</Timestamp>
    <Version versionID="VERMER1" global="false" timestamp="2000-10-01T12:00:00+01:00">
      <ProcessRef>PBASC50t2T1c1C1</ProcessRef>
      <SourceRef>BBAS0</SourceRef>
      <SourceRef>BBAS849</SourceRef>
      <SourceRef>BCDMS0</SourceRef>
      <SourceRef>BCDMS-1921</SourceRef>
      <SourceRef>BCDMS-1681</SourceRef>    </Version>   
    <HomepageUrl>http://www.vamdc.org/activities/research/software/spectcol/</HomepageUrl>
    <Name>Spectcol</Name>
    <Comments>Merging between collisional data from Basecol and spectroscopic data from CDMS by cross matching state with same J values in CO</Comments>
    <Origin xsi:type="VamdcNodeOriginType">  
      <Timestamp>2015-12-03T15:50:21+01:00</Timestamp>
      <Version versionID="VERCDMS1" global="false" timestamp="2000-10-01T12:00:00+01:00">
        <SpeciesRef>XCDMS-83</SpeciesRef>
        <StateRef>SCDMS-83-1</StateRef>
        <StateRef>SCDMS-83-2</StateRef>
        <StateRef>SCDMS-origin-83</StateRef>
        <ProcessRef>PCDMS-R15140649</ProcessRef>
      </Version>   
      <HomepageUrl>http://cdms.ph1.uni-koeln.de/</HomepageUrl>
      <Name>CDMS database</Name>
      <Comments>Spectroscopic data extracted from CDMS VAMDC node.</Comments>
      <Query>select * where (RadTransWavelength &gt;= 2.6006E7 AND RadTransWavelength &lt;= 2.6008E7) AND ((MoleculeStoichiometricFormula = 'CO'))</Query>  
      <OriginIdentifier>ivo://vamdc/cdms/vamdc-tap_12.07</OriginIdentifier>      
    </Origin>
    <Origin xsi:type="VamdcNodeOriginType">
      <Timestamp>2015-12-03T14:40:21+01:00</Timestamp>
      <Version versionID="VER001" global="false" timestamp="2015-09-01T08:10:12+01:00">
        <SpeciesRef>XBAS2</SpeciesRef>
        <StateRef>SBASET54-1</StateRef>        
      </Version>
      <HomepageUrl>http://basecol.vamdc.org</HomepageUrl>
      <Name>Basecol</Name>
      <Comments>Collisional data extracted from Basecol VAMDC node.</Comments>
      <Query>select * where ((target.MoleculeStoichiometricFormula = 'CO')) AND ((collider.AtomSymbol = 'he'))</Query>    
      <OriginIdentifier>ivo://vamdc/basecol/vamdc-tap_12.07</OriginIdentifier>
    </Origin>
  </Origin>
  <Species>
    <Atoms>
      <Atom>
        <ChemicalElement>
          <NuclearCharge>2</NuclearCharge>
          <ElementSymbol>He</ElementSymbol>
        </ChemicalElement>
        <Isotope>
          <IsotopeParameters>
            <MassNumber>4</MassNumber>
            <Mass>
              <Value units="amu">4.0026</Value>
            </Mass>
          </IsotopeParameters>
          <Ion speciesID="XBAS2">
            <IonCharge>0</IonCharge>
            <AtomicState stateID="SBASET54-1">
              <Comments>Energy level of He (no structure)</Comments>
              <SourceRef>BBAS0</SourceRef>
              <AtomicNumericalData>
                <StateEnergy>
                  <Value units="1/cm">0.0</Value>
                </StateEnergy>
              </AtomicNumericalData>
              <AtomicQuantumNumbers>
                <TotalAngularMomentum>0.0</TotalAngularMomentum>
              </AtomicQuantumNumbers>
              <AtomicComposition>
                <Component>
                  <Term>
                    <LS>
                      <L>
                        <Value>0</Value>
                        <Symbol>L</Symbol>
                      </L>
                      <S>0.0</S>
                    </LS>
                  </Term>
                </Component>
              </AtomicComposition>
            </AtomicState>
            <InChIKey>SWQJXJOGLNCZEY-UHFFFAOYSA-N</InChIKey>
          </Ion>
        </Isotope>
      </Atom>
    </Atoms>
    <Molecules>
      <Molecule speciesID="XCDMS-83">
        <MolecularChemicalSpecies>
          <OrdinaryStructuralFormula>
            <Value>CO</Value>
          </OrdinaryStructuralFormula>
          <StoichiometricFormula>CO</StoichiometricFormula>
          <ChemicalName>
            <Value>Carbon Monoxide</Value>
          </ChemicalName>
          <InChI>1S/CO/c1-2</InChI>
          <InChIKey>UGFAIRIUMAVXCW-UHFFFAOYSA-N</InChIKey>
          <VAMDCSpeciesID>UGFAIRIUMAVXCW-UHFFFAOYSA-N</VAMDCSpeciesID>
          <PartitionFunction>
            <T units="K">
              <DataList>1.072 1.148 1.23 1.318 1.413 1.514 1.622 1.738 1.862 1.995 2.138 2.291 2.455 2.63 2.725 2.818 3.02 3.236 3.467 3.715 3.981 4.266 4.571 4.898 5.0 5.248 5.623 6.026 6.457 6.918 7.413 7.943 8.511 9.12 9.375 9.772 10.471 11.22 12.023 12.882 13.804 14.791 15.849 16.982 18.197 18.75 19.498 20.893 22.387 23.988 25.704 27.542 29.512 31.623 33.884 36.308 37.5 38.905 41.687 44.668 47.863 51.286 54.954 58.884 63.096 67.608 72.444 75.0 77.625 83.176 89.125 95.499 102.329 109.648 117.49 125.893 134.896 144.544 150.0 154.882 165.959 177.828 190.546 204.174 218.776 225.0 234.423 251.189 269.153 288.403 300.0 309.03 331.131 354.813 380.189 407.38 436.516 467.735 500.0 501.187 537.032 575.44 616.595 660.693 707.946 758.578 812.831 870.964 933.254 1000.0</DataList>
            </T>
            <Q>
              <DataList>1.01721581835 1.02422837677 1.03341102522 1.04512141245 1.05985020123 1.07774884294 1.09923417933 1.12473229226 1.15441650589 1.18864057708 1.22774305695 1.27174864212 1.32092124733 1.37520246158 1.40530654907 1.43514349042 1.50100121324 1.57270208471 1.65050158216 1.73501239088 1.8265276783 1.92535019369 2.03179140659 2.14652192999 2.18241666936 2.26987115366 2.40252640741 2.54553752733 2.69889427601 2.8632973732 3.04016627563 3.22985295285 3.4334258678 3.65195638949 3.74352703447 3.88615863094 4.13746641418 4.40695607053 4.69606494313 5.00551115749 5.33781506844 5.69369660756 6.0753182363 6.48412210856 6.9226335574 7.12225433182 7.3922960322 7.89599749116 8.4355434648 9.0138233311 9.633727213 10.2977845092 11.0096090065 11.7724538098 12.5895724935 13.4656648032 13.8965040828 14.4043468795 15.4099582553 16.4875620501 17.6425836137 18.8800875183 20.2062236736 21.6271428763 23.1500815655 24.7815541073 26.5302455525 27.4545173427 28.40375754 30.4111396059 32.5625279126 34.8676989471 37.3378779761 39.9850161394 42.8214291179 45.8608826253 49.1175078254 52.6076105394 54.5813579159 56.3475015965 60.3549445917 64.6491571735 69.2508130864 74.1820459087 79.4660945815 81.7185083304 85.1287658147 91.1970158766 97.6997287418 104.669260558 108.868854329 112.139428854 120.146852376 128.732781329 137.942190312 147.825233889 158.438549717 169.84604324 181.685035016 182.121688241 195.35029188 209.62906021 225.070653126 241.803934912 259.976802963 279.755720344 301.328756805 324.90561697 350.71803967 379.021079637</DataList>
            </Q>
          </PartitionFunction>
          <StableMolecularProperties>
            <MolecularWeight>
              <Value units="unitless">28.0</Value>
            </MolecularWeight>
          </StableMolecularProperties>
          <Comment> 28503- v 1:CO; $v=0$</Comment>
        </MolecularChemicalSpecies>
        <MolecularState stateID="SCDMS-83-1">
          <MolecularStateCharacterisation>
            <StateEnergy energyOrigin="SCDMS-origin-83">
              <Value units="1/cm">0.0</Value>
            </StateEnergy>
            <TotalStatisticalWeight>1</TotalStatisticalWeight>
            <NuclearStatisticalWeight>1</NuclearStatisticalWeight>
          </MolecularStateCharacterisation>
          <Case xmlns:xsi="http://www.w3.org/2001/XMLSchema-instance" xsi:type="dcs:Case" caseID="dcs">
            <dcs:QNs>
              <dcs:J>0</dcs:J>
            </dcs:QNs>
          </Case>
        </MolecularState>
        <MolecularState stateID="SCDMS-83-2">
          <MolecularStateCharacterisation>
            <StateEnergy energyOrigin="SCDMS-origin-83">
              <Value units="1/cm">3.845033</Value>
            </StateEnergy>
            <TotalStatisticalWeight>3</TotalStatisticalWeight>
            <NuclearStatisticalWeight>1</NuclearStatisticalWeight>
          </MolecularStateCharacterisation>
          <Case xmlns:xsi="http://www.w3.org/2001/XMLSchema-instance" xsi:type="dcs:Case" caseID="dcs">
            <dcs:QNs>
              <dcs:J>1</dcs:J>
            </dcs:QNs>
          </Case>
        </MolecularState>
        <MolecularState auxillary="true" stateID="SCDMS-origin-83">
          <MolecularStateCharacterisation>
            <StateEnergy energyOrigin="SCDMS-origin-83">
              <Value units="1/cm">0.0</Value>
            </StateEnergy>
            <TotalStatisticalWeight>1</TotalStatisticalWeight>
            <NuclearStatisticalWeight>1</NuclearStatisticalWeight>
          </MolecularStateCharacterisation>
          <Case xmlns:dcs="http://vamdc.org/xml/xsams/1.0/cases/dcs" xsi:type="dcs:Case" caseID="dcs" xsi:schemaLocation="http://vamdc.org/xml/xsams/1.0/cases/dcs ../../cases/dcs.xsd">
            <dcs:QNs>
              <dcs:ElecStateLabel>X</dcs:ElecStateLabel>
              <dcs:v>0</dcs:v>
              <dcs:J>0</dcs:J>
            </dcs:QNs>
          </Case>
        </MolecularState>
      </Molecule>
    </Molecules>
  </Species>
  <Processes>
    <Radiative>
      <RadiativeTransition id="PCDMS-R15140649" process="excitation">
        <SourceRef>BCDMS0</SourceRef>
        <EnergyWavelength>
          <Frequency>
            <Value units="MHz">115271.2021</Value>
            <Accuracy>1.0E-4</Accuracy>
          </Frequency>
        </EnergyWavelength>
        <UpperStateRef>SCDMS-83-2</UpperStateRef>
        <LowerStateRef>SCDMS-83-1</LowerStateRef>
        <SpeciesRef>XCDMS-83</SpeciesRef>
        <Probability>
          <TransitionProbabilityA>
            <Value units="1/cm">7.20360334988E-8</Value>
          </TransitionProbabilityA>
          <IdealisedIntensity>
            <Value units="unitless">-5.0105</Value>
          </IdealisedIntensity>
          <Multipole>E2</Multipole>
        </Probability>
        <ProcessClass>
          <Code>rota</Code>
        </ProcessClass>
      </RadiativeTransition>
    </Radiative>
    <Collisions>
      <CollisionalTransition id="PBASC50t2T1c1C1">
        <Comments> Rotational de-excitation of CO (v=0) by He (Cecchi-Pestellini &amp; al, 2002)</Comments>
        <SourceRef>BBAS0</SourceRef>
        <ProcessClass>
          <Code>inel</Code>
        </ProcessClass>
        <Reactant>
          <SpeciesRef>XCDMS-83</SpeciesRef>
          <StateRef>SCDMS-83-2</StateRef>
        </Reactant>
        <Reactant>
          <SpeciesRef>XBAS2</SpeciesRef>
          <StateRef>SBASET54-1</StateRef>
        </Reactant>
        <Product>
          <SpeciesRef>XCDMS-83</SpeciesRef>
          <StateRef>SCDMS-83-1</StateRef>
        </Product>
        <Product>
          <SpeciesRef>XBAS2</SpeciesRef>
          <StateRef>SBASET54-1</StateRef>
        </Product>
        <DataSets>
          <DataSet dataDescription="rateCoefficient">
            <TabulatedData>
              <Comments>Rate coefficients</Comments>
              <X units="K">
                <DataList>5.0 10.0 20.0 40.0 60.0 80.0 100.0 200.0 300.0 500.0</DataList>
              </X>
              <Y units="cm3/s">
                <DataList>3.4E-11 3.2E-11 3.0E-11 2.8E-11 2.7E-11 2.6E-11 2.6E-11 2.5E-11 2.5E-11 2.6E-11</DataList>
              </Y>
            </TabulatedData>
          </DataSet>
        </DataSets>
      </CollisionalTransition>
    </Collisions>
  </Processes>
  <Sources>
    <Source sourceID="BBAS0">
      <Category>database</Category>
      <SourceName>BASECOL database</SourceName>
      <Year>2015</Year>
      <Authors>
        <Author>
          <Name>M.-L. Dubernet</Name>
        </Author>
      </Authors>
      <UniformResourceIdentifier>http://basecol.obspm.fr</UniformResourceIdentifier>
      <ProductionDate>2015-12-03</ProductionDate>
      <Comments>select * where ((target.MoleculeStoichiometricFormula = 'CO')) AND ((collider.AtomSymbol = 'he'))</Comments>
    </Source>  
    <Source sourceID="BBAS849">
       <Category>journal</Category>
       <SourceName>apj</SourceName>
       <Year>2002</Year>
       <Authors>
         <Author>
           <Name>N. Balakrishnan</Name>
         </Author>
         <Author>
           <Name>A. Dalgarno</Name>
         </Author>
         <Author>
           <Name>C. Cecchi-Pestellini</Name>
         </Author>
         <Author>
           <Name>E. Bodo</Name>
         </Author>
       </Authors>
       <Title>Rotational and Vibrational Excitation of CO Molecules by Collisions with $^{4}$He Atoms</Title>
       <Volume>571</Volume>
       <PageBegin>1015</PageBegin>
       <PageEnd>1020</PageEnd>
       <UniformResourceIdentifier>
    http://adsabs.harvard.edu/cgi-bin/nph-bib_query?
         bibcode=2002JChPh.116.4517K&db_key=PHY
       </UniformResourceIdentifier>
     </Source>
     <Source sourceID="BCDMS0">
    <Comments>This Source is a self-reference.
    It represents the database and the query that produced the xml document.
    The sourceID contains a timestamp.
    The full URL is given in the tag UniformResourceIdentifier but you need
    to unescape ampersands and angle brackets to re-use it.
    Query was:  select * where ((target.MoleculeStoichiometricFormula = 'CO')) AND ((collider.AtomSymbol = 'he'))</Comments>
         <Year>2015</Year>
         <Category>database</Category>
         <UniformResourceIdentifier>http://cdms.ph1.uni-koeln.de/cdms/tap/sync?
         LANG=VSS2&amp;
         REQUEST=doQuery&amp;FORMAT=XSAMS&amp;QUERY= select * where ((target.MoleculeStoichiometricFormula = 'CO')) AND ((collider.AtomSymbol = 'he'))</			UniformResourceIdentifier>
         <ProductionDate>2015-12-03</ProductionDate>
         <Authors>
            <Author>
               <Name>N.N.</Name>
            </Author>
         </Authors>
      </Source>
      <Source sourceID="BCDMS-1921">
         <Authors>
            <Author>
               <Name>Müller, H. S. P.</Name>
            </Author>
            <Author>
               <Name>Endres, C. P.</Name>
            </Author>
            <Author>
               <Name>Schlemmer, S.</Name>
            </Author>
            <Author>
               <Name>Stutzki, J.</Name>
            </Author>
         </Authors>
         <Title />
         <Category>database</Category>
         <Year>2012</Year>
         <SourceName>CDMS database</SourceName>
      </Source>
      <Source sourceID="BCDMS-1681">
         <Authors>
            <Author>
               <Name>Winnewisser, G.</Name>
            </Author>
            <Author>
               <Name>Belov, S. P.</Name>
            </Author>
            <Author>
               <Name>Klaus, T.</Name>
            </Author>
            <Author>
               <Name>Schieder, R.</Name>
            </Author>
         </Authors>
         <Title />
         <Category>journal</Category>
         <Year>1997</Year>
         <SourceName>J. Mol. Spectrosc.</SourceName>
         <Volume>184</Volume>
         <PageBegin>468</PageBegin>
         <DigitalObjectIdentifier>10.1006/jmsp.1997.7341</DigitalObjectIdentifier>
      </Source>
  </Sources>
  <Comments>Data merged by SPECTCOL.</Comments>
</XSAMSData>
\end{lstlisting}


\bibliographystyle{plain}
\bibliography{biblio}

\end{document}